**Introducing a Symmetry-Breaking Coupler into a Dielectric Metasurface Enables Robust High-Q Quasibound States in the Continuum and Efficient Nonlinear Frequency Conversion**

*Gianni Q. Moretti[1], Andreas Tittl[2], Emiliano Cortés[2], Stefan A. Maier[2,3,4], Andrea V. Bragas[1], Gustavo Grinblat[1*]*

[1]Departamento de Física, FCEN, IFIBA-CONICET, Universidad de Buenos Aires, C1428EGA Buenos Aires, Argentina

[2]Chair in Hybrid Nanosystems, Nanoinstitute Munich, Faculty of Physics, Ludwig-Maximilians-Universität München, 80539 München, Germany.

[3]School of Physics and Astronomy, Monash University, Clayton Victoria 3800, Australia

[4]Department of Physics, Imperial College London, London SW7 2AZ, UK

*Email: grinblat@df.uba.ar

**Abstract.** Dielectric metasurfaces supporting quasi-bound states in the continuum (quasi-BICs) exhibit very high quality factor resonances and electric field confinement. However, accessing the high-$Q$ end of the quasi-BIC regime usually requires marginally distorting the metasurface design from a BIC condition, pushing the needed nanoscale fabrication precision to the limit. This work introduces a novel concept for generating high-$Q$ quasi-BICs, which strongly relaxes this requirement by incorporating a relatively large perturbative element close to high-symmetry points of an undistorted BIC metasurface, acting as a coupler to the radiation continuum. We validate this approach by adding a ~100 nm diameter cylinder between two reflection-symmetry points separated by a 300 nm gap in an elliptical disk metasurface unit cell, using gallium phosphide as the dielectric. We find that high-$Q$ resonances emerge when the cylindrical coupler is placed at any position between such symmetry points. We further explore this metasurface's second harmonic generation capability in the optical range. Displacing the coupler as much as a full diameter from a BIC condition produces record-breaking normalized conversion efficiencies >$10^2$ W$^{-1}$. The strategy of enclosing a disruptive element between multiple high-symmetry points in a BIC metasurface could be applied to construct robust high-$Q$ quasi-BICs in many geometrical designs.

## 1. Introduction

Dielectric metasurfaces can greatly enhance incident electromagnetic fields over a large area with very low absorption in the visible and infrared regions of the spectrum. They can controllably tailor the amplitude and phase of the light wavefront,[1–4] efficiently generate nonlinear light,[5,6] and can be exploited for biosensing[7–9] and metrology applications,[10,11] among others. Recently, resonances originating from the concept of bound states in the continuum (BICs) have been shown to produce quality factors and field enhancements far higher than those of widely studied radiative Mie modes and nonradiative anapole states.[12–14] BICs are perfectly confined states which possess an infinite lifetime and therefore cannot couple to radiation channels.[15] Symmetry-protected BICs, the most studied type of BIC,[16] can be realized by introducing a symmetry-breaking perturbation, which typically involves slightly distorting the metasurface geometrical design, turning the optically inaccessible BIC into a high quality factor quasi-BIC. Some examples in the literature include unit cells of pairs of slightly differently sized rods[17,18] or cylinders,[19,20] blocks[21] and rings[22] with missing sections, as well as tilted elliptical cylinders.[23,24] The magnitude of the asymmetry in relation to the characteristic dimensions of the system can often be quantified by a dimensionless parameter α; for sufficiently small perturbations, a proportionality relationship between the quality factor (*Q*, defined as the resonance frequency divided by its linewidth) and $\alpha^{-2}$ can be obtained for most metasurface designs.[25]

Given the excellent field confinement abilities of quasi-BICs and the large intrinsic nonlinearities of high-index dielectrics, these platforms are especially useful to enhance nonlinear processes at sub-wavelength volumes. Over the past decade, second and third harmonic generation (SHG and THG, respectively) have been extensively studied using metallic and dielectric nanostructures,[26–30] with research only recently focusing on quasi-BIC resonances.[17,24,31–34] Obtained experimental conversion efficiencies ($\eta_{SHG} = P_{2\omega}/P_\omega$ and $\eta_{THG} = P_{3\omega}/P_\omega$) reached ~0.01%,[32,35,36] while normalized efficiencies ($\xi_{SHG} = \eta_{SHG}/P_\omega$ and $\xi_{THG} = \eta_{THG}/P_\omega^2$) up to 0.04 W$^{-1}$ (SHG)[24] and $1 \times 10^{-5}$ W$^{-2}$ (THG)[21] were achieved. On the other hand, several theoretical works have predicted significantly better performances.[6,18,37,38] However, as they have considered the perturbative nonlinear regime, their validity only holds for sufficiently small incident intensities at relatively low field amplifications. Because of the strong near-field enhancements present in quasi-BICs, pump depletion and the optical Kerr effect become important. This issue has been recently addressed for SHG, demonstrating a maximum theoretical conversion efficiency of 0.2% at 100 kW/cm$^2$ in an AlGaAs metasurface.[39] However, an unrealistic asymmetry parameter as small as $\alpha < 0.02$ was assumed, well below the current experimental limit of $\alpha \sim 0.1$ for the same system design.[24]

In this work, we propose a different method to break symmetries in BIC-based metasurfaces. Instead of slightly modifying the geometry of the meta-atoms, we add a perturbative element to the unaltered BIC metasurface to enable coupling to radiation channels. By placing such an element close to a relevant high-symmetry point of the array, the resulting metasurface supports very high-*Q* resonances. As a unit cell can hold more than one such symmetry point, the dependence of the quality factor on the introduced asymmetry is, in general, no longer monotonous.

We focus on the extensively studied BIC metasurface design defined by a unit cell of two elliptical cylinders.[7,8,13,24] To generate the quasi-BIC, the reflection symmetry of this system is usually broken by slightly tilting the elliptical disks toward each other. Here, we show that a resonance of the same nature is more efficiently excited by adding an off-center cylindrical *coupler* while keeping the elliptical disks parallel (see schematic in Figure 1). We choose gallium phosphide (GaP) as the high-index dielectric (n > 3), as it possesses negligible absorption in the majority of the visible spectrum and large nonlinearities.[40–42] We model the SHG capabilities of our design in the nonperturbative regime and demonstrate conversion efficiencies up to 0.5% at pump intensities <1 kW/cm$^2$.

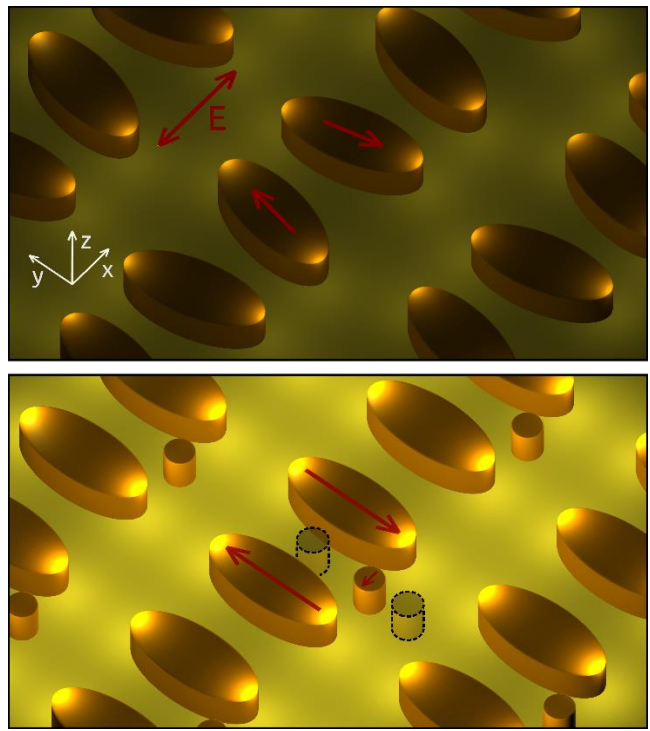

**Figure 1. Elliptical cylinder metasurfaces with introduced reflection asymmetries in the *y*-direction.** For the configuration on the top, an *x*-polarized electric field excites a quasi-BIC resonance, as enabled by the inward tilting of the meta-atoms. A more intense quasi-BIC (bottom) of the same character can be excited by adding an off-center cylindrical coupler between reflection-symmetry positions, marked with semi-transparent cylinders in a representative unit cell.

## 2. Results

The geometries of the studied GaP metasurfaces are depicted in Figure 2A. The short ($D_1$) and long diameters ($D_2$) of the elliptical cylinders are 170 nm and 380 nm, respectively, the height is 200 nm, the center-to-center distance between adjacent meta-atoms ($d$) is 300 nm, and the square lattice period ($p$) is 600 nm. For the conventional metasurface configuration (left panel), used as a reference, the asymmetry parameter is defined as $\alpha_1 = \sin(\theta)$, with $\theta$ the tilt angle. In the proposed design (right panel), $\theta$ is kept fixed at $\theta = 0$, and a cylinder of diameter $D$ is added between two reflection-symmetry points (marked as crosses) separated by $w = 300$ nm, acting as a coupler to radiation channels. In this case, the asymmetry parameter is defined as $\alpha_2 = \Delta y/w$, where $\Delta y$ is the displaced distance of the cylindrical coupler in the $y$-direction from the bottom reflection-symmetry point. Note that the vertical reflection symmetry is broken in both metasurfaces, while the horizontal reflection symmetry is maintained.

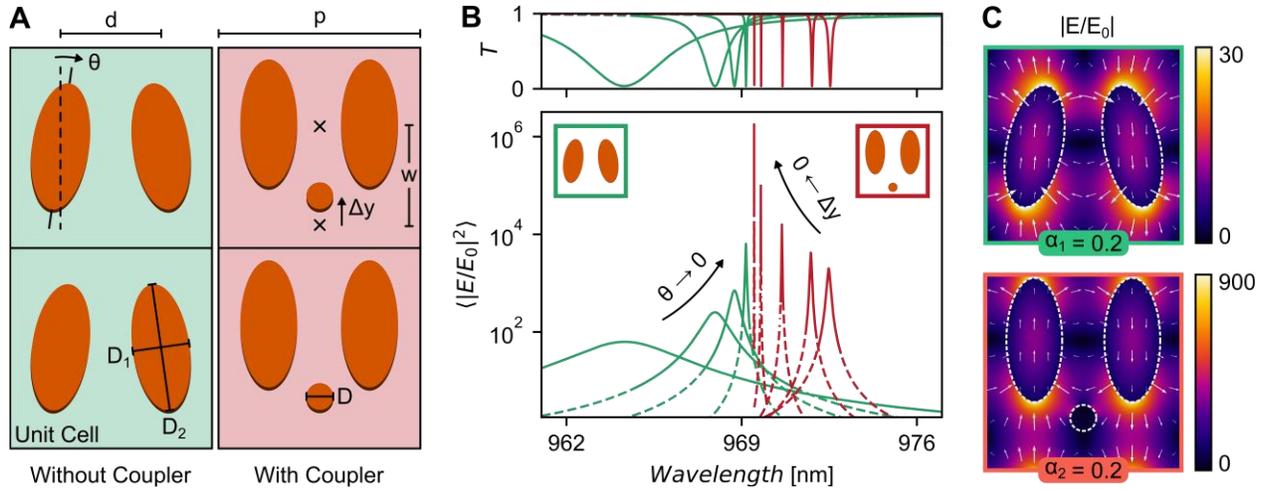

**Figure 2. Field enhancement performance of quasi-BIC metasurfaces. (A)** Schematics of the analyzed metasurfaces, showing two adjacent unit cells for each of them. On the left, the asymmetry is introduced by tilting the meta-atoms inward. On the right, a cylindrical coupler is added between two reflection-symmetry points. **(B)** Computed transmission ($T$) of the metasurfaces (top) and average intensity enhancement inside the dielectric (bottom) as a function of incident wavelength. Different $\theta$ and $\Delta y$ values are compared to illustrate the behavior when approaching zero asymmetry. **(C)** Electric field distribution (magnitude and direction) in the $xy$ plane when considering the same asymmetry parameter value for both metasurfaces.

To compare the optical resonances of the two configurations, we initially set the coupler diameter to $D = 80$ nm and compute the transmission ($T$) and the average intensity enhancement inside the dielectric ($\langle |E/E_0|^2 \rangle$) as a function of the incident wavelength for $\alpha$ approaching 0 (see Methods for specifics on numerical simulations). The tilt angle is varied from 10 to 1° ($\alpha_1$ ranging from ~0.2 to ~0.02) and $\Delta y$ from 150 to 10 nm ($\alpha_2$ from 0.5 to ~0.03), respectively. As can be seen in Figure 2B, the quality factor and the field enhancement diverge when $\theta \rightarrow 0$ and $\Delta y \rightarrow 0$, while the resonance position converges to (approximately) the same BIC wavelength from opposite sides of the spectrum. However, significantly larger field enhancements are found for the three-element unit cell design. To better illustrate the advantage of this geometry, Figure 2C compares the electric field distribution for both metasurfaces at $\alpha_{1,2} = 0.2$ (corresponding to $\theta = 12°$ and $\Delta y = 60$ nm, respectively). We observe that the proposed design provides a 30-fold increase in field amplification compared to the reference case, which corresponds to an increase of the intensity enhancement by a factor of ~1000. In both metasurfaces, the excited mode consists of in-plane electric quadrupole and out-of-plane magnetic dipole components, which couple to the external $x$-polarized incident radiation through an electric dipole in the $x$-direction, as enabled by the introduced asymmetries[13,43] (details of the multipolar decompositions can be found in Figure S1 of the Supporting Information). Note that making $\Delta x \neq 0$ (i.e., displacing the coupler in the $x$-direction) at $\Delta y \neq 0$ would introduce additional asymmetry, decreasing the quality factor of the resonance, as shown in Figure 3A for $\Delta y = 40$ nm. On the other hand, taking $\Delta x \neq 0$ at $\Delta y = 0$ would produce no quasi-BIC resonance, since it would unbalance the compensated electric field of the BIC in the $y$-direction only, preventing any coupling with the incoming $x$-polarized beam, as can be deduced from the ideal BIC field distribution in Figure 3B. Breaking the symmetry solely via $\Delta y \neq 0$ as in Figure 2 corresponds to moving the coupler along a $E_y = 0$ line (see Figure 3B, right panel), enabling fine control of the electric polarization in the *x*-direction, with no further distortion of the BIC condition.

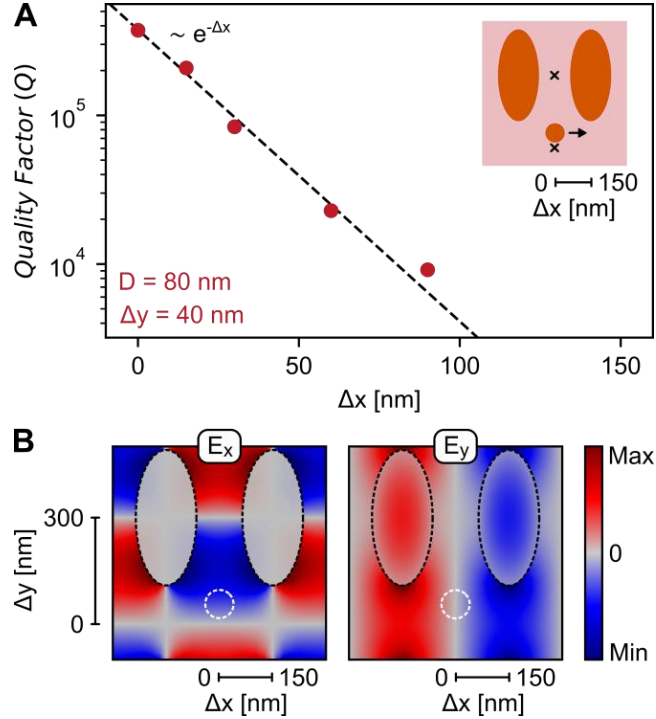

**Figure 3. Breaking the reflection symmetry in the *x*-direction.** **(A)** Quality factor of the resonance when displacing the cylindrical coupler along the *x*-direction at $\Delta y = 40$ nm. A nearly exponential decay is observed when increasing $\Delta x$. **(B)** In-plane electric field distribution of the ideal BIC condition (parallel elliptical cylinders with no coupler element). The eigenstate has fully compensated in-plane electric fields, with no possible coupling to linearly polarized light. Adding a coupler to this configuration (as represented in white) would unbalance the BIC field distribution depending on its position.

To evaluate the influence of the coupler diameter on the resonant properties, in Figure 4A we present the dependence of $Q$ on $D$ as a function of $\alpha_2$ at $\Delta x = 0$. We find that the proportionality relationship between $Q$ and $\alpha_2^{-2}$ holds for all explored diameters for sufficiently small displacement asymmetries. We also find that enlarging the size of the coupler at same $\alpha_2$ decreases the quality factor of the resonance, acting as an alternate way of increasing the degree of symmetry perturbation. Nevertheless, even for disk diameters as large as 160 nm, this system performs significantly better than the reference metasurface (shown as a solid line). This can be understood by noting that tilting the elliptical disks can be thought of as displacing four bulky couplers (the four tips of the two meta-atoms) from the BIC condition, corresponding to a much larger symmetry distortion.

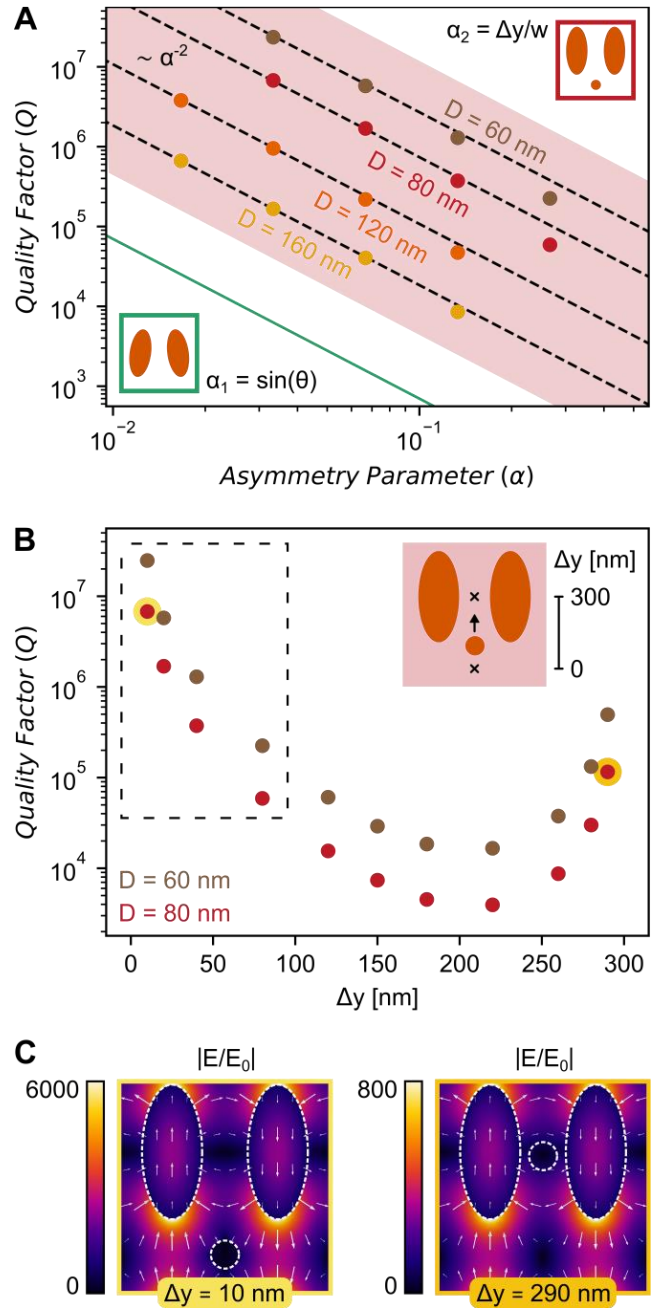

**Figure 4. Detailed analysis of the quasi-BIC resonances. (A)** Quality factor of the resonance as a function of the asymmetry parameter for both configurations, considering different coupler diameters. The dotted and solid lines follow a linear relationship between $Q$ and $\alpha^{-2}$. Maximum $\alpha_2$ in the plot corresponds to $\Delta y$ = 80 nm. **(B)** Quality factor for $\Delta y$ values up to 290 nm, for two particular coupler diameters. The dashed region corresponds to the asymmetries displayed in (A). **(C)** Electric field distribution close to the two BIC conditions for $D$ = 80 nm.

As can be seen in Figure 2 and Figure 3, the proposed metasurface presents two high-symmetry points with $E_x = 0$ and $E_y = 0$, corresponding to $\Delta y = 0$ and $\Delta y = 300$ nm, connected by a $E_y = 0$ line. Hence, when continuously varying $\Delta y$, a second BIC condition should be found. Furthermore, since $E_x$ varies with a steeper slope near $\Delta y = 300$ nm, compared with $\Delta y = 0$ nm (see left panel of Figure 3B), the resulting stronger perturbation of the near-fields should produce weaker quasi-BIC resonances. This becomes evident when plotting the dependence of $Q$ on $\Delta y$ for the whole displacement range, as shown in Figure 4B for two disk diameters, $D = 60$ nm and $D = 80$ nm, which are sizes that can fit comfortably within the 130-nm gap between adjacent elliptical cylinders. In contrast to the reference metasurface, the quality factor never approaches zero, staying always above $10^3$ and $10^4$ for $D = 80$ nm and $D = 60$ nm, respectively, as the coupler is "trapped" between two BIC conditions. As shown in Figure 4C, the nature of the excited quasi-BIC remains the same when approaching both high-symmetry points, with reduced field enhancement near the top reflection-symmetry point as anticipated. These findings offer great advantages for the realization of BIC-based metasurfaces with high robustness against fabrication inaccuracies, as even experimentally straightforward displacement distances of 100 nm can produce quality factors as high as $10^5$. In comparison, the reference design would require an unfeasible tilt angle of <0.5°, which corresponds to moving the tips of the meta-atoms from their straight positions by just ~1 nm. While the coupler can be conveniently chosen to be of any size and placed at positions weakly perturbing the BIC field distribution, the conventional strategy of tilting the meta-atoms displaces large pieces of material of fixed size around their immediate surrounding, strongly disturbing high-field regions.

The enhanced electric field inside the metasurface makes it attractive for studying nonlinear processes arising from the intrinsic nonlinear properties of GaP. In Figure 5, we analyze the SHG performance of our metasurface, considering a diameter of 80 nm for the cylindrical coupler (see

Methods for nonlinear simulation details). First, we study how the nonlinear response varies when modifying the in-plane orientation of GaP crystal lattice. The [100] crystal direction is set initially along the *x*-direction of the coordinate system. As seen in Figure 5A, a rotation in the crystal plane of $\beta = 45°$ gives the maximum enhancement of efficiency. This corresponds to the [110] crystal direction becoming aligned with the fundamental electric field inside the material, which accommodates mainly along the *y*-direction (see Figure 2C and Figure 4C). The nonlinear polarization is most efficiently excited in this condition given the symmetry of GaP nonlinear susceptibility tensor.[44]

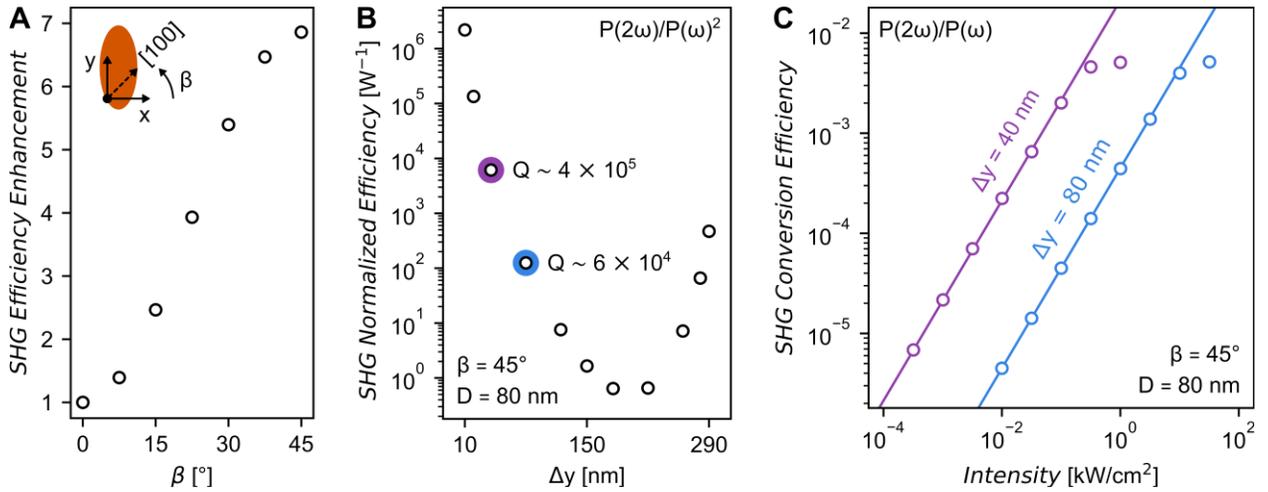

**Figure 5. Second harmonic efficiency calculations. (A)** SHG efficiency enhancement when varying the in-plane crystal orientation angle $\beta$. **(B)** Normalized conversion efficiency at $\beta = 45°$ as a function of $\Delta y$ for $D = 80$ nm, computed in the perturbative regime. **(C)** Dependence of the SHG conversion efficiency on the pump intensity for the two cases highlighted in (B), considering pump depletion and the optical Kerr effect. The computed data (shown as circles) deviate from the perturbative behavior (solid lines) at approximately 0.1 and 10 kW/cm$^2$, respectively.

In Figure 5B, the normalized SHG efficiency in the perturbative regime is calculated as a function of $\Delta y$ in the 10-290 nm range at $\beta = 45°$, revealing values between ~1 and $10^6$ W$^{-1}$. In particular, for metasurfaces with quality factors around $10^5$ ($\Delta y \sim$ 40-80 nm), above which nonradiative contributions due to sample imperfections - not considered in our model - become important,[21] the normalized efficiency reaches the order of $10^3$ W$^{-1}$. This performance exceeds the highest previously reported theoretical values by three orders of magnitude.[14] We then study the nonperturbative regime, considering pump depletion and the optical Kerr effect, which are no longer negligible even for relatively low pump intensities. In this regime, the resulting large second harmonic field interacts with and depletes the fundamental field, while the ultra-high field enhancement distorts the resonant condition through the intensity-dependent refractive index (to expand on this refer to Figure S3 in the Supporting Information). Consequently, we find that the conversion efficiency saturates around 0.5%, as can be seen in Figure 5C. We also observe that the saturation power decreases when increasing the field confinement ability of the resonator (i.e., for smaller asymmetries), as expected. Remarkably, for the metasurface with $\Delta y = 40$ nm, the saturation regime would be reached with pump powers as low as 1 mW, when considering an excitation area of hundreds of $\mu m^2$ covering thousands of unit cells. These conditions could be achieved with a simple low-cost continuous-wave laser pointer and a common low numerical aperture lens. We note that the damage threshold of GaP in its transparency range is about 0.7 GW/cm$^2$,[45] which is well above the effective pump intensity of <0.1 GW/cm$^2$ that we compute by considering the field enhancement within the dielectric material.

## 3. Conclusion

In summary, we demonstrate a novel strategy to generate robust high-$Q$ quasi-BICs in dielectric metasurfaces. By introducing a symmetry-breaking element into a BIC-supporting unit cell, the coupling to the radiation continuum can be controlled with much better precision than through the usual distorting of the actual metasurface. We test this approach by adding a small cylinder (60-160 nm diameter) to an array of parallel elliptical cylinder GaP meta-atoms. By enclosing the cylindrical coupler between two BIC reflection-symmetry points, high-$Q$ resonances emerge along the whole 300-nm long region between the BIC conditions. We then optimize the SHG capabilities of this system by choosing the more convenient in-plane GaP crystal orientation and perform perturbative and nonperturbative nonlinear studies, considering pump depletion and the optical Kerr effect. In the perturbative regime, the conversion efficiency grows linearly with the intensity enhancement of the resonance, while a maximum conversion efficiency of 0.5% is obtained in the nonperturbative approach, at only <1 kW/cm$^2$ pump intensity. The ultra-strong nonlinear response of this metasurface makes it a very suitable candidate for various nonlinear nanophotonic applications. Furthermore, the robustness of the proposed method for breaking symmetries in BIC metasurfaces can also benefit many other unit cell configurations and improve the performance of a broad range of practical applications from ultrasensitive biodetection to surface-enhanced energy conversion processes.

## 4. Methods

The linear and nonlinear numerical calculations were performed using the RF module of the COMSOL Multiphysics® software.[46] The GaP structures composing the unit cell were placed on a glass/air interface in a square prism domain geometry with periodic boundary conditions in the four lateral faces and a perfectly-matched layer (PML) in the top and bottom faces.

The linear optical studies were carried out by solving the problem for the scattered field, using the analytical expressions of the reflection and transmission Fresnel coefficients at the air/glass interface to define the background field. The complex permittivity of GaP used for the simulations was taken from the literature.[40] Values of $\varepsilon = 2.25$ and $\varepsilon = 1$ for the glass and air domains were utilized, respectively.

The nonlinear calculations were performed with an iterative, segregated approach. For the perturbative simulation, the resulting linearly excited fields within the nanostructures were used to compute a second-order polarization with frequency $2\omega$ in the material, then used as a source to obtain the nonlinear fields. In the nonperturbative condition (pump depletion and optical Kerr effect added to the linear polarization), the second harmonic solution was used again in the linear computation, leading to a different nonlinear field. This was iteratively done until convergence was achieved.

The second-order susceptibility tensor used to compute the nonlinear polarization was taken from the literature.[41,42] GaP has a zinc-blende crystal structure with a nonlinear tensor of the form $\chi^{(2)}_{ijk} \neq 0$ for $i \neq j \neq k$, with all six non-zero components having the same value. We used $\chi^{(2)}$ values ranging from $4 \times 10^{-10}$ to $10^{-10}$ mV$^{-1}$ in the 800-1200 nm wavelength range. For the optical Kerr effect, we computed a third-order susceptibility of $4.3 \times 10^{-19}$ m$^2$V$^{-2}$, derived from the experimentally reported magnitude of the nonlinear index $n_2 \sim 10^{-17}$ m$^2$W$^{-1}$.[42] With these considerations, the added polarizations to the field equations at $\omega$ and $2\omega$ are:

$$P_i(\omega) = \varepsilon_0 \chi^{(2)} E_{2j} E^*_{1k} + \varepsilon_0 \chi^{(3)}(|E_1|^2 E^*_{1i} + 2E_{1i}(E_{1j}E^*_{1j} + E_{1k}E^*_{1k}))$$

$$P_i(2\omega) = 2\varepsilon_0 \chi^{(2)} E_{1j} E_{1k}$$

where, in all cases, $i \neq j \neq k$, $E_1 = E(\omega)$ and $E_2 = E(2\omega)$.

The SHG conversion efficiency was computed as the ratio between the nonlinear Poynting vector integrated over the top and bottom limits of the simulation domain and the incident power over the surface area of a unit cell.

The multipolar decompositions shown in the Supporting Information were calculated using cartesian decomposition.[47]

**Acknowledgments**

This work was partially supported by PICT 2017-2534, PICT 2019-01886, PIP 112 201301 00619, PIP 112 202001 01465, UBACyT Proyecto 20020170100432BA, and UBACyT Proyecto 20020190200296BA. We also acknowledge funding and support from the Deutsche Forschungsgemeinschaft (DFG, German Research Foundation) under Germany's Excellence Strategy – EXC 2089/1 – 390776260, the Bavarian program Solar Energies Go Hybrid (SolTech), the Center for NanoScience (CeNS), and the Lee-Lucas Chair in Physics.

**Conflict of Interest**

The authors declare no conflict of interest.

**Data Availability Statement**

The data supporting this study's findings are available from the corresponding author upon reasonable request.

**Introducing a Symmetry-Breaking Coupler into a Dielectric Metasurface Enables Robust High-Q Quasibound States in the Continuum and Efficient Nonlinear Frequency Conversion**

Gianni Q. Moretti[1], Andreas Tittl[2], Emiliano Cortés[2], Stefan A. Maier[2,3,4], Andrea V. Bragas[1], Gustavo Grinblat[1*]

[1]Departamento de Física, FCEN, IFIBA-CONICET, Universidad de Buenos Aires, C1428EGA Buenos Aires, Argentina

[2]Chair in Hybrid Nanosystems, Nanoinstitute Munich, Faculty of Physics, Ludwig-Maximilians-Universität München, 80539 München, Germany.

[3]School of Physics and Astronomy, Monash University, Clayton Victoria 3800, Australia

[4]Department of Physics, Imperial College London, London SW7 2AZ, UK

*Email: grinblat@df.uba.ar

**Contents**

Figure S1: Multipolar decomposition of the quasi-BIC resonance.

Figure S2: Additional linear results.

Figure S3: Additional SHG results.

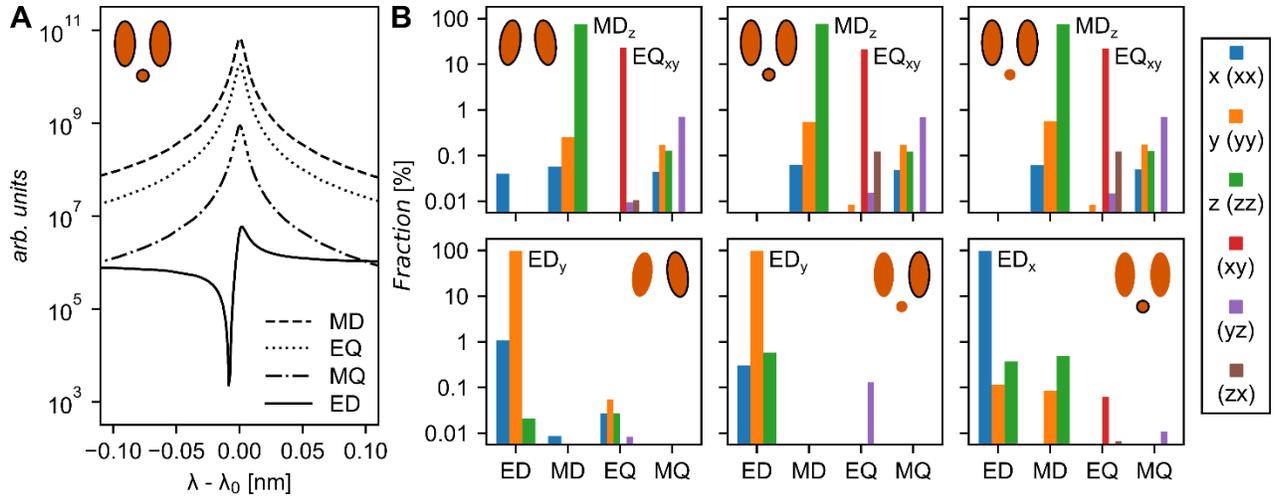

**Figure S1: Multipolar decomposition of the quasi-BIC resonance. (A)** Electric dipole (ED), magnetic dipole (MD), electric quadrupole (EQ) and magnetic quadrupole (MQ) moments of the system with the coupler as a function of incident wavelength (D = 80 nm, $\Delta y$ = 60 nm). **(B)** Cartesian fractions of all contributions for different integration volumes (marked as black outlines) for the two metasurfaces ($\theta = 1°$, reference metasurface; D = 80 nm and $\Delta y$ = 60 nm, proposed design). The largest contributions (>10%) are labeled in each case. When integrating over all meta-atoms (first row, two graphs on the left), the out-of-plane MD and the in-plane EQ are the most significant components. As they cannot be excited by linearly polarized light, a coupling mechanism is needed. The *x*-polarized ED appears by tilting the elliptical disks in the reference metasurface, while no ED is present in them for the proposed design. Instead, it appears in the cylindrical coupler, as can be seen in the last panel of the bottom row. The first two panels in the bottom row show that each elliptical cylinder has an ED in the *y*-direction, which cancels out when considering the full dimer.

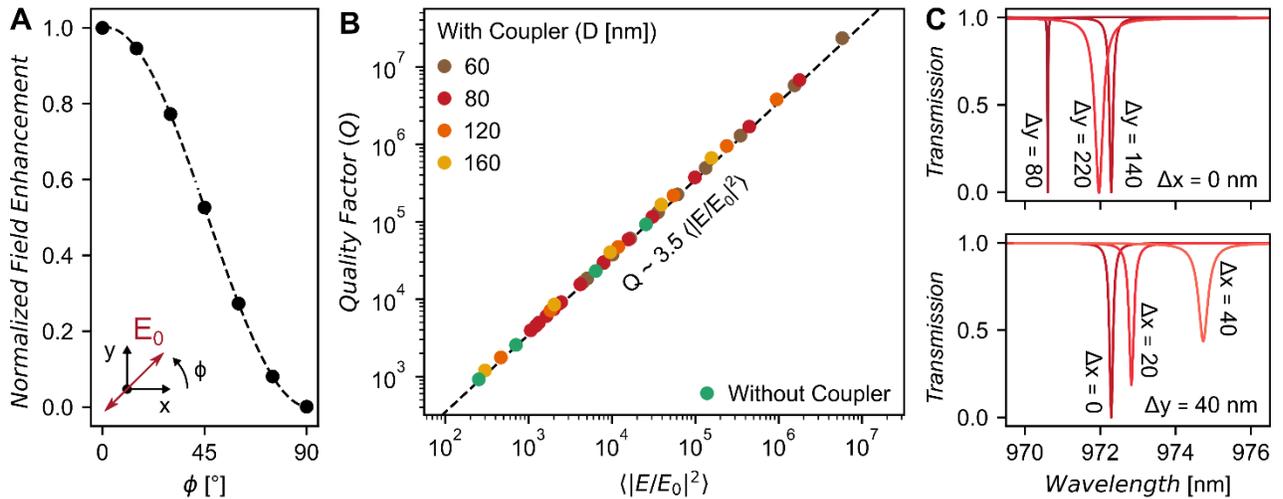

**Figure S2: Additional linear results. (A)** Enhancement of the electric field as a function of the polarization of the incident light, showing that the resonance is most efficiently excited with *x*-polarized light. **(B)** Quality factor against averaged intensity enhancement inside the material. A linear proportionality relation exists for all resonances studied in Figure 2 and Figure 4, main text. **(C)** Computed transmission of the metasurface for different $\Delta x$ and $\Delta y$ for D = 80 nm. Because of the imposed geometry and polarization conditions, the transmission does not reach zero for $\Delta x \neq 0$.

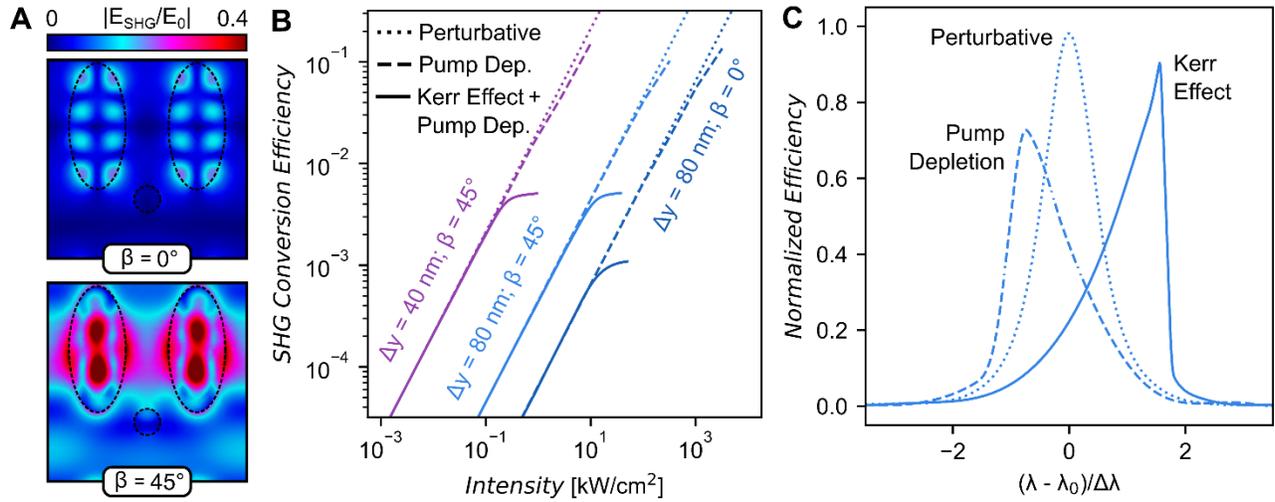

**Figure S3: Additional SHG results. (A)** Second harmonic field for two in-plane GaP crystal orientations. The maximum nonlinear enhancement occurs when the [110] crystal direction aligns with the *y*-direction. **(B)** Dependence of the SHG conversion efficiency on the pump intensity for the cases highlighted in Figure 5B (main text), also including a case with $\beta = 0°$, when considering different nonlinear regimes (perturbative, pump depletion, and Kerr effect + pump depletion). Note that the pump depletion term becomes relevant for intensities far higher compared to the Kerr effect. It can be observed that modifying the crystal orientation does not change the saturation intensity but does change the maximum conversion efficiency. **(C)** Nonlinear response in the saturation regions as a function of wavelength ($\lambda$), measured relative to the perturbative resonance value ($\lambda_0$) and normalized by peak width ($\Delta\lambda$). The resonance shifts to the blue when having only pump depletion and to the red when adding the optical Kerr effect.